\documentclass[baaa]{baaa}

\usepackage[pdftex]{hyperref}
\usepackage{subfigure}
\usepackage{natbib}
\usepackage{helvet,soul}
\usepackage[font=small]{caption}

\defcitealias{PlanckCollaboration2014}{Planck Collaboration et al. (2014)}

\contriblanguage{1}

\contribtype{1}

\thematicarea{6}

\title{Enhanced sub-resolution star formation \\
models in cosmological simulations}

\titlerunning{Enhanced star formation models in cosmological simulations}

\author{
E. Lozano\inst{1}, 
C. Scannapieco\inst{1,2}, 
S.E. Nuza\inst{3,2}, 
Y. Ascasibar\inst{4,5},
L. Biaus\inst{1} \&
F.G. Iza\inst{3}
}

\authorrunning{Lozano et al.}

\contact{elozano@df.uba.ar}

\institute{
Departamento de Física, Facultad de Ciencias Exactas y Naturales, UBA, Argentina \and Consejo Nacional de Investigaciones Científicas y Técnicas, Argentina \and
Instituto de Astronomía y Física del Espacio, CONICET--UBA, Argentina \and
Departamento de Física Teórica, Campus de Cantoblanco, UAM, Madrid, Spain \and Centro de Investigación Avanzada en Física Fundamental (CIAFF-UAM), 28049 Madrid, Spain
}

\resumen{
Uno de los componentes cruciales para simular el crecimiento y la evolución de galaxias en un marco cosmológico es el modelizado de la formación estelar (SF) y su correspondiente retroalimentación. Tradicionalmente, la ley de SF se basa en la relación empírica de Kennicutt-Schmidt, una correlación entre la tasa de formación estelar (SFR) y la densidad total de gas. Recientemente, se ha encontrado que la SFR se correlaciona fuertemente con la abundancia de hidrógeno molecular ($\mathrm{H}_2$). Esto da lugar a la pregunta de si el $\mathrm{H}_2$ es un precursor necesario para la SF o es simplemente un trazador de la densidad total ya que ambos escenarios explicarían las correlaciones observadas. En este trabajo estudiamos el impacto de una ley de SF basada en el $\mathrm{H}_2$ en simulaciones cosmológicas de la formación de galaxias de tipo Vía Láctea. Nuestro modelo predice la formación de una galaxia de disco que, comparada con la receta tradicional, retrasa la formación estelar por aproximadamente $500 \, \mathrm{Myr}$, lo que resulta en una menor SFR, menor tamaño del disco y una proporción mayor de gas neutro a ionizado. Estos hallazgos resaltan la importancia de incluir modelos de SF sofisticados, que puedan ser contrastados con distintas observaciones -- incluidas aquellas relacionadas al $\mathrm{H}_2$ -- para lograr un mejor entendimiento de los procesos involucrados en la evolución galáctica.
}

\abstract{
One of the crucial components in simulating the growth and evolution of galaxies within a cosmological framework is the modeling of star formation (SF) and its corresponding feedback. Traditionally, the implemented SF law follows the empirical Kennicutt-Schmidt relation, which links the SF rate (SFR) in a gas element to the total gas density. More recently, an even stronger correlation has been observed between the SFR and the amount of molecular hydrogen ($\mathrm{H}_2$). This opens up the question of whether molecular hydrogen is a necessary precursor for SF or is instead a tracer of the total amount of gas, both of which would explain the observed correlations. In this study, we examine the impact of using an $\mathrm{H}_2$-based SF law on the formation of Milky Way-mass galaxies using cosmological, hydrodynamical simulations. We find that the galaxy modeled with our approach exhibits a well-defined, disc-like morphology. Compared to the traditional recipe, our model delays the onset of SF by approximately $500 \, \mathrm{Myr}$ resulting in a lower SFR, a smaller disc size, and a higher proportion of neutral to ionized gas within the disc region. These findings highlight the importance of including sophisticated SF models which can be compared to several observations -- including those related to $\mathrm{H}_2$ -- to better understand the processes affecting galaxy formation and evolution.
}


\begin{document}

\maketitle

\section{Introduction}\label{sec:introduction}

Within the framework of the standard cosmological model, galaxy formation occurs as gas condenses within dark matter halos. These halos emerge from the gravitational collapse of small density fluctuations and grow over time through the accretion of matter and smaller substructures. From the earliest stages of galaxy formation, various physical processes work together to shape the properties of galaxies, forming a complex interplay of mechanisms that act throughout cosmic history.

A key component in simulations of galaxy formation and evolution is the accurate modeling of star formation (SF) and its associated feedback. The process of star formation begins with gas cooling, which determines the amount and distribution of dense gas available for SF. The formation of stars, in turn, enriches the interstellar medium (ISM) with chemical elements and energy. The interplay between gas cooling and the heating caused by supernova explosions is central in regulating SF rates (SFRs) across time and space. However, because star formation and stellar feedback occur on smaller, unresolved scales, they must be included through sub-grid models that account for their influence on larger scales.

Over the past decades, observational studies have revealed that the correlation between gas density and SFR becomes even stronger when molecular gas is considered (\citealt{Leroy2008,Bigiel2010,Bolatto2011} and references therein). This relationship holds both on resolved scales \citep{Baker2021} and at galaxy-wide scales across various redshifts \citep{Baker2022}. Moreover, the typical timescale of star formation appears to depend on the physical properties of the ISM \citep{Bigiel2008,Bigiel2010,Bolatto2011,Leroy2013}. These findings have inspired the development of more advanced star formation models within simulations and semi-analytic frameworks \citep{Murante2010,Molla2015,Millan-Irigoyen2020}.

In this study, we investigate the impact of one such molecular hydrogen-based star formation law on galaxy formation. Our approach is based on the semi-analytical model developed by \citet{Millan-Irigoyen2020}, which assumes a multiphase ISM composed of atomic, ionized, and molecular gas, as well as metals and dust. This model is implemented within the moving-mesh, magnetohydrodynamical code \texttt{AREPO} \citep{Springel2010}, which represents the gas component as a collection of discrete cells. Each cell is assumed to have a multiphase structure, including ionized, atomic, and molecular hydrogen phases, along with a stellar component. Metals are also tracked using the standard chemical evolution model of \texttt{AREPO}. The mass exchange between the phases is described by a set of differential equations that account for the effects of various physical processes. Star formation within the stellar phase is linked to the molecular hydrogen fraction and is governed by a density-dependent timescale.

This proceeding is structured as follows. Section~\ref{sec:simulations} outlines the simulation code, the implemented physics, and the newly developed star formation model. In Section~\ref{sec:results}, we utilize our model to study the formation of a galaxy with a Milky Way-like mass within a cosmological framework and analyze the predicted properties of the simulated system, with a focus on the different phases, as well as the star formation activity. Lastly, Section~\ref{sec:conclusions} presents our conclusions.

\section{The simulation and star formation model}\label{sec:simulations}

\subsection{The \texttt{AREPO} code}
\label{arepo}

We utilize the $N$-body magnetohydrodynamics (MHD) code \texttt{AREPO} to carry out our simulations. In our implementation, we only modify the SF prescription assigned to each gas cell, while leaving the rest of the code -- including the feedback, cooling, and metal enrichment -- unchanged. In the standard SF implementation, star particles are stochastically generated from gas cells that meet the necessary conditions for SF, and the SFR is proportional to the total gas density of the cell. In our model, instead, the SFR depends on integrating four differential equations, as we explain below. For a detailed description of the main \texttt{AREPO} code, we refer the reader to \cite{Springel2010}.

To simulate the formation and evolution of a Milky Way-mass galaxy, we use the initial conditions (ICs) of the Au6 halo (level 4 resolution) from the Auriga Project \citep{Grand2017}. These ICs correspond to a {\it zoom-in} halo region with a virial mass of $M_{200} = 1.05 \times 10^{12} \, \mathrm{M_\odot}$ at $z = 0$, which is relatively isolated. In the original Auriga simulations, these ICs produce a galaxy with properties similar to the Milky Way, as discussed in \cite{Grand2017}. Throughout this proceeding, we compare the predictions of two simulations: Au6\_MOL adopting our new SF recipe, and Au6\_STD that uses the standard SF recipe (Eq.~\ref{eq:sfr_arepo}).  

The cosmological parameters adopted in this study follow those from \citetalias{PlanckCollaboration2014}: $\Omega_{\rm M} = 0.307$, $\Omega_{\rm b} = 0.048$, $\Omega_\Lambda = 0.693$, and a Hubble constant $H_0 = 100 \, h \, \mathrm{km \, s^{-1} \, Mpc^{-1}}$, with $h = 0.6777$. All other input parameters for the code are identical to those used in the Auriga Project \citep{Grand2017}.  

\subsection{The star formation model}\label{subsec:model_equations}

In our star formation model, star particles are formed from multi-phase gas cells using a stochastic approach. This method, described in \cite{Springel2003} (see also \citealt{Grand2017}), allows gas cells with densities above a threshold value ($\rho_\mathrm{th}$) to be eligible for star formation. In the standard implementation, star particles are stochastically formed at a rate defined by  
\begin{equation} 
    \label{eq:sfr_arepo}
    \dot{m}_* = \frac{x \, M_\mathrm{cell}}{t_*(\rho)} \, ,  
\end{equation}  
where $x$ represents the fraction of cold gas, $M_\mathrm{cell}$ is the mass of the gas cell, and $t_*(\rho)$ is the star formation timescale, given by  
\begin{equation}  
    t_* = t_0^* \left(\frac{\rho}{\rho_\mathrm{th}}\right)^{\!-1/2} \, ,  
\end{equation}  
with $t_0^*$, the maximum star formation timescale, being a parameter of the model. With this choice, the SFR density is proportional to the $3/2$ power of the gas density, as in the Kennicutt-Schmidt law.

In our approach, instead of using Eq.~\ref{eq:sfr_arepo}, we calculate the SFR for each gas cell by solving a set of differential equations. These equations track the evolution of the mass fractions of four components: (1) an ionized phase at a temperature of approximately $\sim \! 10^4 \, \mathrm{K}$, (2) an atomic phase at around $\sim \! 100 \, \mathrm{K}$, (3) a molecular phase at $\sim \! 10 \, \mathrm{K}$, and (4) stars\footnote{Although the gas is composed of hydrogen, helium, and metals, the model only includes the dominant hydrogen reactions that exchange mass between phases.}.
Mass exchange between phases occurs through several processes: recombination of ionized gas, condensation of atomic hydrogen, photodissociation of molecular gas, photoionization of atoms, star formation, and mass return from stars to the interstellar gas. We represent mass fractions of each phase relative to the total cell mass as $f_X = M_X / M_\mathrm{cell}$, where $X$ can be $i$ (ionized), $a$ (atomic), $m$ (molecular), or $s$ (stellar). At any time $t$, the total mass of a gas cell is:  
\begin{equation}  
    M_\mathrm{cell}(t) = M_i(t) + M_a(t) + M_m(t) + M_s(t) \, ,
\end{equation}  
where the sum of the corresponding mass fractions must equal unity, i.e. 
\begin{equation}  
    f_i + f_a + f_m + f_s = 1 \, .  
\end{equation}  

\begin{figure}[ht]
    \centering  
    \includegraphics[width=\columnwidth]{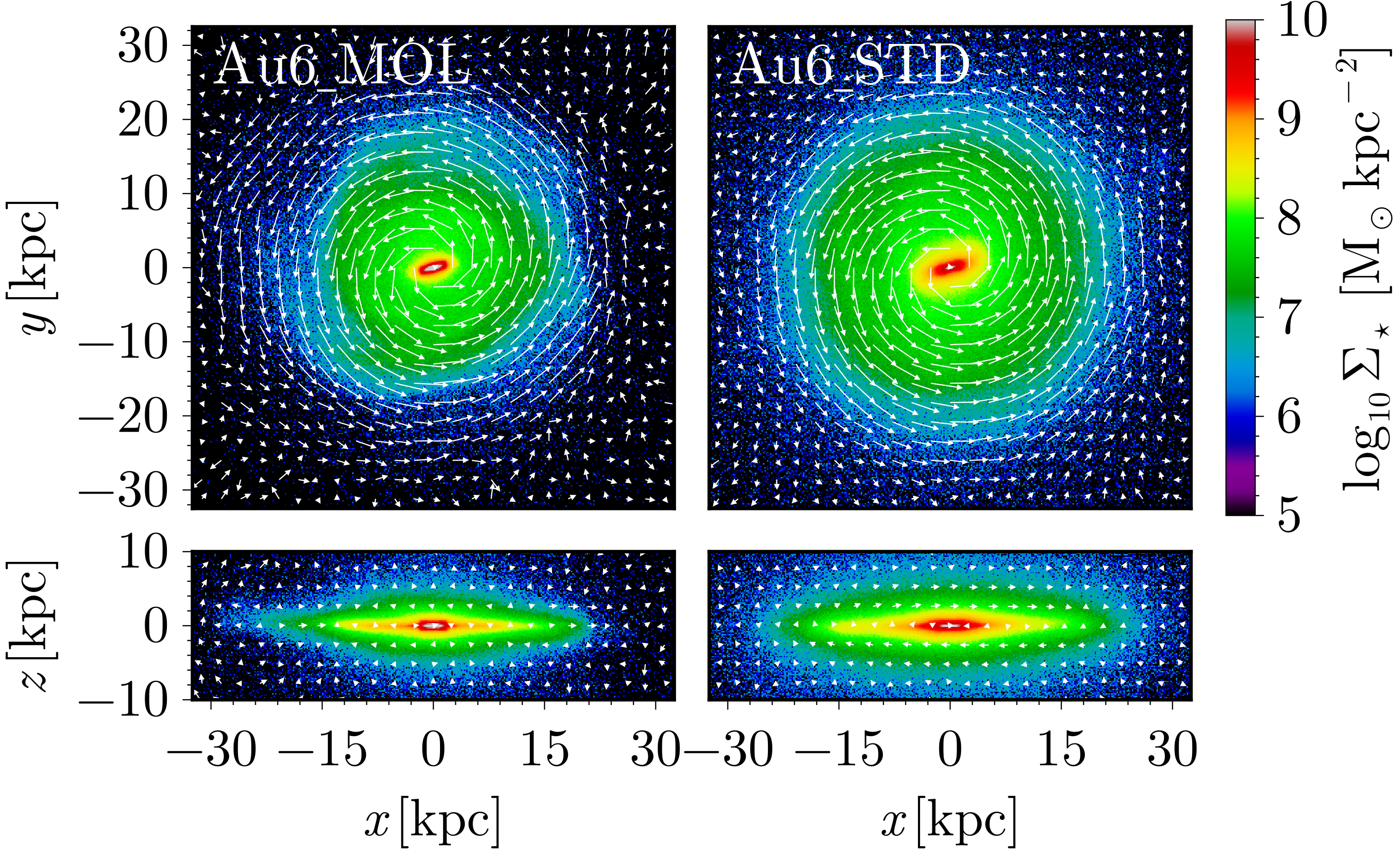}  
    \caption{Projected stellar mass distributions (face-on and edge-on views) for Au6\_MOL and Au6\_STD at $z = 0$. The white arrows indicate the corresponding velocity field.}  
    \label{fig:stellar_maps_Au6}  
\end{figure}  

The evolution of these phases is governed by the following set of equations:  
\begin{subequations}  
    \label{eqs:odes}  
    \begin{align}  
        \frac{d}{dt}f_i(t) &= - \frac{f_i(t)}{\tau_r(t)} + \eta_i \, \psi(t) + R \, \psi(t) \, , \\  
        \frac{d}{dt}f_a(t) &= \frac{f_i(t)}{\tau_r(t)} - \eta_i \, \psi(t) - \frac{f_a(t)}{\tau_c(t)} + \eta_d \, \psi(t) \, , \\  
        \frac{d}{dt}f_m(t) &= \frac{f_a(t)}{\tau_c(t)} - \eta_d \, \psi(t) - \psi(t) \, , \\  
        \frac{d}{dt}f_s(t) &= \left(1 - R\right)\psi(t)\, ,  
    \end{align}  
\end{subequations}  
where $\psi(t)$ represents the fractional instantaneous SFR. We assume that $\psi(t)$ depends solely on the amount of molecular gas and is regulated by the star formation timescale $\tau_s$. Specifically, following \citet{Millan-Irigoyen2020} we take 
\begin{equation}  
    \label{eq:sfr}  
    \frac{d}{dt} f_s(t) \biggr\rvert_{\mathrm{SFR}} =: \psi(t) = \frac{f_m}{\tau_s} \, .
\end{equation}  
The mass exchange due to recombination, condensation, and star formation is modeled using timescales ($\tau_r$, $\tau_c$, and $\tau_s$, respectively), which depend on the properties of the gas and vary with time and cell conditions. Meanwhile, the process of photoionization, photodissociation, and mass return are assumed to be proportional to $\psi(t)$, with fixed proportionality factors ($\eta_i$, $\eta_d$, and $R$, respectively). These factors are computed semianalytically and only depend on metallicity, as shown in \citet{Millan-Irigoyen2020}.

We solve the system of Eqs.~\ref{eqs:odes} for each gas cell eligible for star formation\footnote{ 
Cells entering the star formation routine for the first time are assumed to have no stellar or molecular mass, with their initial ionized and atomic fractions determined by the standard treatment of ionized and neutral phases in \texttt{AREPO}.}, integrating over the time step under the assumption that the mass, density, and metallicity of the gas cell remain constant during the integration. This allows us to estimate the SFR per gas cell as the mass in the stellar phase divided by the total integration time ($t_i$), i.e.:
\begin{equation}  
    \label{eq:sfr_mol}
    \dot{m}_* = \frac{f_s \, M_\mathrm{cell}}{t_i} \, , 
\end{equation}   
which we use to calculate the probability of a gas cell to stochastically form a star particle. 

\begin{figure}[ht]
    \centering  
    \includegraphics[width=0.8\columnwidth]{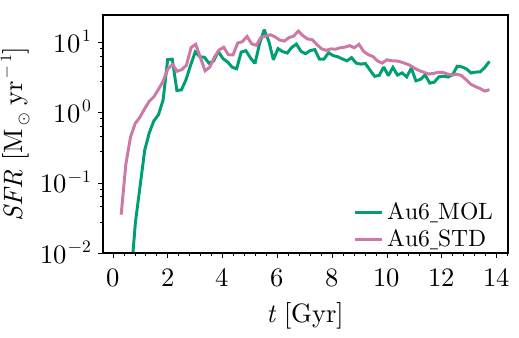}  
    \caption{The SFR history derived from stellar ages at $z = 0$ for Au6\_MOL and Au6\_STD.}  
    \label{fig:sfr_Au6}  
\end{figure}  

\section{Results}
\label{sec:results}

\subsection{Properties of the simulated galaxies at $z = 0$}

Fig.~\ref{fig:stellar_maps_Au6} shows the projected stellar mass distribution of Au6\_MOL and Au6\_STD at $z = 0$, both face-on and edge-on (with the $z$-axis aligned to the stellar angular momentum). Both galaxies exhibit a spiral-like morphology with an extended, rotationally supported disc and a central spheroidal component. The velocity field of Au6\_MOL suggests that the stellar disc exhibits ordered rotation up to a radius of approximately $20 \, \mathrm{kpc}$, while Au6\_STD extends up to $\sim 30 \, \mathrm{kpc}$.

Fig.~\ref{fig:sfr_Au6} shows the SFRs for both simulations. Au6\_MOL shows a rapid early growth during the first few Gyr, stabilizing at $\sim \! 4 \, \mathrm{M_\odot \, yr^{-1}}$ after $\sim \! 3 \, \mathrm{Gyr}$, but with a delay of $\sim 500 \, \mathrm{Myr}$ compared to Au6\_STD. Throughout the whole evolution, Au6\_STD has a consistently higher SFR than Au6\_MOL, although they end up with similar values at $z = 0$. Our findings confirm that stellar feedback plays a fundamental role in regulating the SFRs, although differences in the evolution might be important to decide on the most realistic model.

\subsection{Gas components at $z = 0$}
\label{sec:z0_gas}

We have measured the disc sizes for the various components using the radius enclosing $95\%$ of the total mass \citep{Iza2022}\footnote{In the case of the gas, we define a maximum radius of $40 \, \mathrm{kpc}$, given the significant fraction of gas beyond the disc region.}. We find that all gas components have comparable sizes of $\sim 33 \, \mathrm{kpc}$ for Au6\_MOL, and are slightly higher $\sim 38 \, \mathrm{kpc}$, for Au6\_STD. These results are consistent with our model producing a less extended stellar disc, as shown in Fig.~\ref{fig:stellar_maps_Au6}.

\begin{figure}[ht]
    \centering  
    \includegraphics[width=0.75\columnwidth]{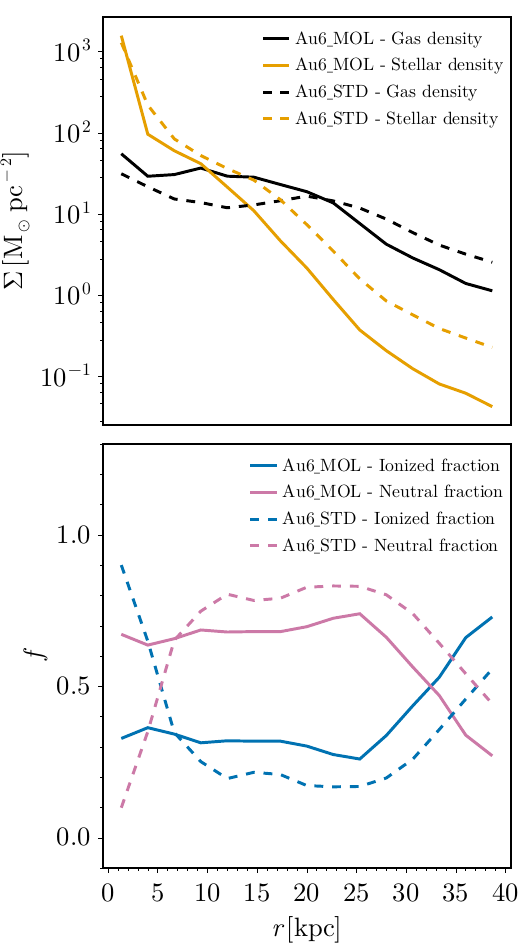}  
    \caption{{\it Top panel}: Projected surface density profiles for the stars and gas of Au6\_MOL and Au6\_STD at $z = 0$. {\it Bottom panel}: Radial mass fraction profiles of the ionized and neutral phases, normalized to the total gas mass.}  
    \label{fig:density_profile_gas_Au6}  
\end{figure} 

Fig.~\ref{fig:density_profile_gas_Au6} shows the radial profiles of the stellar and gas components for Au6\_MOL and Au6\_STD, at $z = 0$. The gas density profiles are similar for both simulations but, consistently with our previous findings, the stellar profile has higher values in Au6\_STD compared to Au6\_MOL. This is a consequence of the different conversion rates of gas to stars in the two models. The bottom panel of this figure shows the radial profiles, at $z = 0$, for the ionized and neutral components. We find that the neutral gas in Au6\_MOL dominates over the ionized component in the whole disc region, unlike in Au6\_STD where we find the opposite relation in the inner $5 \, \mathrm{kpc}$. These results might be used to, via detailed comparisons with available observations, help disentangle the effects of using $\mathrm{H}_2$ as a necessary precursor for SF on the properties of the interstellar medium.

\section{Conclusions}\label{sec:conclusions}

In this work, we introduced a model to describe the ionized, atomic, and molecular components of hydrogen in numerical simulations of galaxy formation. We also analyzed the implications of linking star formation to the molecular phase, aiming to provide a more realistic framework for star formation.  

Our approach builds on the formalism presented in \cite{Millan-Irigoyen2020}, adapting it to work with the chemical and feedback modules of the magnetohydrodynamical, cosmological simulation code \texttt{AREPO}. Each gas cell is modeled as a mixture of ionized, atomic, and molecular hydrogen, alongside stars. Mass exchange between these components occur through several processes: recombination (ionized to atomic), condensation (atomic to molecular), star formation (molecular to stellar), photodissociation (molecular to atomic), photoionization (atomic to ionized), and mass return (stellar to ionized). While the processes are modeled in a simple way, they incorporate the main variations in efficiencies and timescales, which depend on the properties of the ISM.

To test our model, we simulated the formation of a Milky Way-mass galaxy using a cosmological {\it zoom-in} initial condition from the Auriga Project \citep{Grand2017} and compared our star formation model (Au6\_MOL) to the standard \texttt{AREPO} implementation (Au6\_STD), which ties star formation to the total gas mass. Run Au6\_MOL showed a lower SFR during the whole evolution, resulting at $z = 0$ in a galaxy with a smaller stellar disc and a higher neutral gas fraction than Au6\_STD, while maintaining similar structural properties reminiscent of a spiral galaxy with a young, rotationally supported stellar disc and a small bulge.

This work represents our first attempt to model molecular hydrogen formation in cosmological simulations. While simple enough to simulate individual galaxies in a cosmological setting (effectively avoids the complexities of ISM physics on $\lesssim 100 \, \mathrm{pc}$ scales), it still captures the relation between the ISM and the SF process at the resolved scales.

\begin{acknowledgement}
S.E.N. and C.S. are members of the Carrera del Investigador Científico of CONICET. They acknowledge support from Agencia Nacional de Promoción Científica y Tecnológica through PICT 2021-GRF-TI-00290 and Universidad de Buenos Aires through UBACyT 20020170100129BA. 
\end{acknowledgement}


\bibliographystyle{baaa}
\small
\bibliography{bibliografia}
 
\end{document}